\documentclass[12pt,preprint]{aastex}
\usepackage{psfrag}
\usepackage[]{natbib}
\usepackage{multirow}
\begin{document}

\title{Evidence for Cosmic Ray Acceleration in Cassiopeia A}

\author{Miguel Araya\altaffilmark{1} and Wei Cui\altaffilmark{2}}
\affil{Department of Physics, Purdue University,
    West Lafayette, IN 47906, USA}

\altaffiltext{1}{marayaar@purdue.edu}
\altaffiltext{2}{cui@purdue.edu}

\slugcomment{Submitted to ApJ on 5/10/2010}

\begin{abstract}
Combining archival data taken at radio and infrared wavelengths with state-of-the-art measurements at X-ray
and gamma-ray energies, we assembled a broadband spectral energy distribution (SED) of Cas A, a young supernova
remnant. Except for strong thermal emission at infrared and X-ray wavelengths, the SED is dominated by non-thermal
radiation. We attempted to model the non-thermal SED with a two-zone leptonic model which assumes that the radio
emission is produced by electrons that are uniformly distributed throughout the remnant while the
non-thermal X-ray emission is produced by electrons that are localized in regions near the forward shock. Synchrotron
emission from the electrons can account for data from radio to X-ray wavelengths. Much of the GeV--TeV emission
can also be explained by a combination of bremsstrahlung emission and inverse Compton scattering of
infrared photons.
However, the model cannot fit a distinct feature at GeV energies. This feature can be well accounted for
by adding a $\pi^{0}$ emission component to the model, providing evidence for cosmic ray production in Cas A.
We discuss the implications of these results.
\end{abstract}

\keywords{ISM: individual objects (Cassiopeia A), - ISM: supernova remnants, - radiation mechanisms: non-thermal}

\section{Introduction}
Ever since their discovery, cosmic rays have been studied extensively. They consist of mostly protons and helium ions (but also heavy
nuclei) and reach energies up to $10^{21}$ eV. At energies below $\sim 10^{15}$ eV, cosmic rays are thought to be of Galactic origin.

The most promising sources of Galactic cosmic rays are supernova remnants (SNRs). They are not only energetically favorable but also provide a mechanism
to accelerate particles to very high energies. The process involves diffusion and scattering of particles across the shock front. The diffusive shock
acceleration naturally leads to a power-law distribution of particles, which is observed in cosmic rays, and with an index that depends only on
the compression factor of the shock \citep{bel78,bla87}.

The link between SNRs and high-energy electrons was initially supported by radio observations that showed synchrotron emission from these objects. The
\emph{Chandra X-Ray Observatory} opened a new era of high-resolution X-ray imaging of SNRs. There is clear evidence for the
existence of higher-energy electrons within these objects, which usually produce thin X-ray ``filaments'' associated with shocks
\citep[e.g.,][]{got01,hwa02,lon03,rho02}. The X-ray spectra of the filaments are consistent with synchrotron emission from particles as they move in an amplified magnetic field $\,\,\,\,\,\,\,\,\,\,$ \citep[e.g.,][]{ber02,vink03,ber04,ara10}.

Despite all the findings that link SNRs with high-energy electrons, direct evidence for the acceleration of ions by these objects is still lacking.
In order to find a signature for hadronic acceleration, studying the photon spectrum of the sources at gamma-ray energies becomes fundamental. It is
expected that at least a fraction of the emission at high energies would be caused by the decay of neutral pions, which should be produced through collisions
of accelerated hadrons with cold ambient protons and ions. Other mechanisms can also contribute to the gamma-ray spectrum besides hadronic interactions.
The high-energy electrons that are responsible for the synchrotron emission at radio and X-ray wavelengths can upscatter ambient photons to TeV energies;
and interactions between these electrons and other charged particles result in bremsstrahlung emission, which can also contribute to the emission in the GeV -- TeV range. It then becomes important not only to obtain observations in the gamma-ray regime but also to model the broadband leptonic emission in detail, in order to separate and quantify the hadronic contribution.

An example of a young SNR where high-quality broadband data are available is Cas A. The data should allow detailed studies of the properties of
high-energy electrons in this remnant. For instance, radio images of Cas A indicate that the synchrotron emission is spread out within its shell,
implying that the electrons responsible for it are likewise more or less uniformly distributed. The X-ray emitting electrons are mainly confined near the
forward shock, with their displacement being strongly limited by synchrotron loss~\citep{got01}.

In this work, we present evidence for cosmic ray acceleration in Cas A from modeling its SED over a broad spectral range (from radio to gamma ray).

\section{Data}
\subsection{Archival data}
In order to construct a broadband SED, we used results from the literature, except for the GeV band where we analyzed public data from the LAT instrument on the
\emph{Fermi} satellite. The radio fluxes were measured at the DKR-1000 radio telescope of the Pushchino Radio Astronomy Observatory \citep{vit65,art67} and at the Staraya Pustyn'
Radio Astronomy Observatory and taken from Vinyaikin \cite{vin06,vin07}, at the frequencies 38, 151.5, 290, and 927 MHz for the epoch 2005.5. An
additional Very Large Array (VLA)\footnote{The VLA array is an instrument of the National Radio Astronomy Observatory, a facility of the
National Science Foundation operated under cooperative agreement by Associated Universities, Inc.} flux at 74 MHz corresponding to the epoch 2005.2
was included \citep{hel09}. The VLA images of the remnant\footnote{See the National Radio Astronomy Observatory image archive, http://images.nrao.edu/395}
show that the radio emission is more diffuse, compared to the X-ray emission.

The infrared fluxes were taken from the \emph{Infrared Astronomical Satellite} (\emph{IRAS}) survey from 12 $\mu$m to 100 $\mu$m \citep{dwe87}. They are likely dominated by thermal emission from dust. Although we are only interested in non-thermal emission from Cas A, the infrared emission may be an important source of seed photons for leptonic gamma ray production.

The X-ray fluxes were obtained from deep exposures of Cas A with the \emph{Chandra X-ray Observatory} ~\cite{hwa04}. Data reduction was carried out as described in Araya et al. (2010). We excluded events from the central compact object in deriving the X-ray spectrum of the
remnant. The X-ray spectrum of the remnant is dominated by thermal emission from highly ionized
ejecta \cite{bor96,fav97}, as evidenced by the presence of numerous lines.  Non-thermal X-ray
emission manifests itself in thin filaments associated with the forward shock, as well as in the hard
tail of the X-ray spectrum.

The TeV gamma-ray measurements were taken from Acciari et al. (2010). They are based on observations of Cas A with VERITAS.

\subsection{Analysis of \emph{Fermi} LAT data}
We analyzed the \emph{Fermi} LAT data that were obtained between 2008 August 4 and 2010 March 10. For details of the instrument, see Atwood et al. \cite{atw09}.

The data analysis was performed with the LAT Science Tools package \footnote{See http://fermi.gsfc.nasa.gov/ssc}. The cuts and selection criteria
applied to the data were those recommended by the Fermi Science Support Center. We selected diffuse events by excluding events with the lowest probability of being gamma rays with the {\em gtselect} tool and used the current version of the instrument response functions given by $P6\_V3\_DIFFUSE$
\cite{ran09}. The energy of the selected events ranges from 200 MeV to 300 GeV. We made appropriate time selections with the {\em gtmktime} tool. Specifically, we used a value
of $105\,^{\circ}$ for the zenith-angle cut to exclude times of high background that is associated with gamma rays produced in the Earth atmosphere.

Due to the large LAT point-spread function at low energies, it becomes necessary to include events within a region around the source. This region of interest is taken here as a $10\,^{\circ}$ radius circle around the cataloged position of Cas A. The accuracy of the analysis is improved when the contributions from other sources within a larger region (``source region'') are also included. We used a $20\,^{\circ}$ radius
source region, which is recommended for sources near the Galactic plane such as Cas A. The background included point sources at fixed positions located in the source region, based on the recently released 11 month Fermi LAT Source Catalog\cite{abd10a} \footnote{See also http://fermi.gsfc.nasa.gov/ssc/data/access/lat/1yr\_catalog/},
as well as the galactic diffuse component (as specified in {\em gll\_iem\_v02.fit}) and the isotropic extragalactic emission (as specified in {\em isotropic\_iem\_v02.txt}).

\begin{figure}[htp]
\centering
\includegraphics[width=12cm,height=10cm]{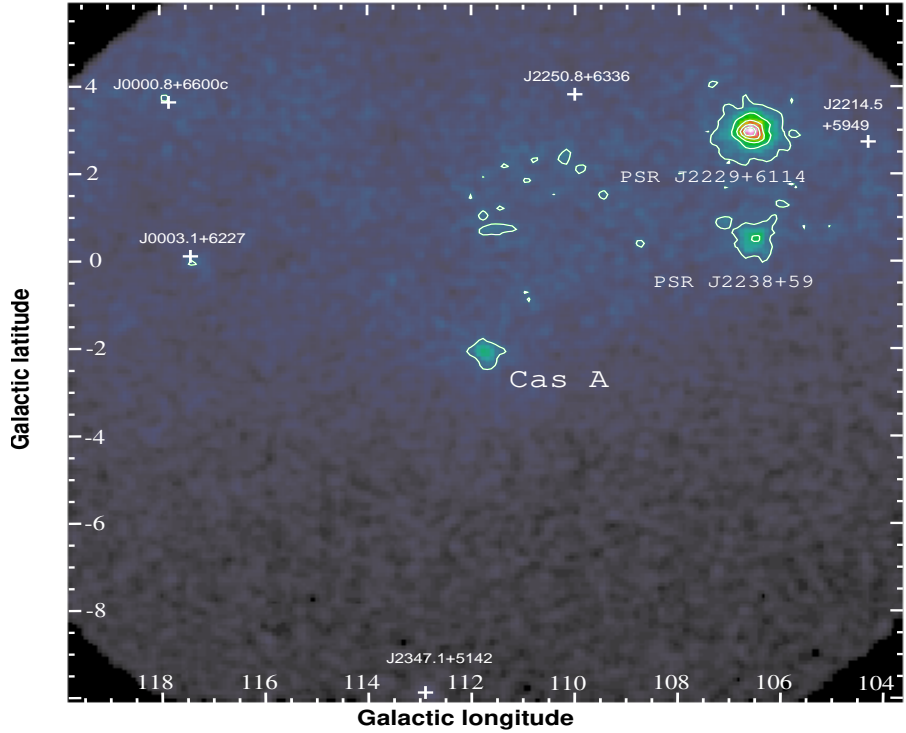}
\caption[Cas A's LAT count map]{Smoothed \emph{Fermi} LAT count map (0.5-60 GeV) of a $16\, ^{\circ} \times 16 \, ^{\circ}$ region around Cas A.
A binning of square pixels $0^{\circ}.08$ in size was used and smoothing was done (in ds9) with a Gaussian kernel
of 2 pixel width. Contours are shown at the levels of 14, 28, 43, and 57 counts. The positions of additional sources found in the \emph{Fermi} Catalogue
and included in the LAT data analysis are also indicated.
\label{fig-1}}
\end{figure}

Figure \ref{fig-1} shows a LAT image of the region near Cas A. A maximum likelihood method (as implemented in the {\em gtlike} tool) is usually used to determine spectral parameters describing LAT sources of interest and their statistical significances~\cite{mat96}. We found that Cas A was detected only at energies above 500 MeV. Between 500 MeV and 60 GeV, it was detected at a significance of $20.3\,\sigma$. The best-fit position of the source, calculated with the {\em gtfindsrc} tool, was
R.A. (J2000)= 23$^{\mbox{h}}$23$^{\mbox{m}}$21.9$^{\mbox{s}}$, decl. (J2000)= 58$^\circ$49$'$47$''$.4, with a statistical uncertainty circle of radius $0^{\circ}.033$ at the $90\%$ confidence level. This position is in agreement with that derived by Abdo et al. (2010b). It also lies inside the VERITAS error box \citep{acc10}.

The \emph{Fermi} LAT spectrum of the source can be fitted with a power-law, $dN/dE \,\, \propto \,\, E^{-p}$, with a photon index of $p=2.06 \pm 0.07$. The
measured 0.5--60 GeV flux is $(1.1 \pm 0.1) \times 10^{-8}$ photons cm$^{-2}$ s$^{-1}$. Note that only statistical errors are shown here. The systematic
errors on flux measurements have been estimated to be 5\% at 500 MeV and 20\% at 10 GeV \citep{ran09}. Our results are in overall agreement with those
reported by the \emph{Fermi} LAT team based on a smaller data set \citep{abd10b}.

\section{Theoretical Modeling}
Figure \ref{fig-2} shows the broadband SED of Cas A. At infrared wavelengths, the SED is dominated by thermal emission from dust (which lies much above the extrapolated radio spectrum). Similarly, at X-ray wavelengths, the thermal emission from hot plasma stands out as the main component, as evidenced by the presence of numerous lines. In this work, we do not attempt to model the thermal emission, but instead focus on the radiation of non-thermal origin.

At radio frequencies the spectrum of the source is of power-law shape, characteristic of optically thin synchrotron emission by high-energy electrons whose spectral energy distribution is also a power law. The existence of non-thermal electrons at even higher energies has been most clearly reaffirmed by observations obtained with the \emph{Chandra X-ray Observatory}, which reveal the presence of thin compact regions of non-thermal X-ray emission~\citep{got01}. To provide a measure of the contribution of non-thermal emission to the overall X-ray spectrum, we also show, in Figure~\ref{fig-2}, the scaled-up X-ray spectrum of a representative non-thermal filament. Note that the X-ray spectrum is quite similar among non-thermal filaments near the forward shock (Araya et al. 2010).

To account for the observed radio and X-ray emission, we constructed a model that involves two populations of electrons: diffuse electrons that are assumed to be uniformly distributed over the entire SNR and localized electrons that are spatially concentrated near the forward shock. The synchrotron radiation from the former is expected to contribute mainly to radio and IR fluxes (but could also reach X-ray wavelengths if there are enough particles of sufficient energy) while that from the latter mainly to X-ray fluxes. Both populations may contribute to the observed gamma-ray fluxes via inverse Compton (IC) scattering and bremsstrahlung processes.

\subsection{Leptonic emission}
The two populations of electrons are modeled as zone 1 consisting of spherical ``blobs'' that approximate the non-thermal filaments seen, and zone 2 consisting of electrons uniformly distributed over a sphere of radius $R_{\mbox{SNR}} = 2.5$ pc, assuming a distance $d_{\mbox{SNR}}=3.4$ kpc \cite{ree95}.

In reality, the radio emission shows spatial structures. Detailed spectral properties of the source such as a flattening of the spectrum above
$\nu \sim 30\,$GHz have been explained by Atoyan et al. (2000), who developed a two-zone model for the radio emission consisting of a diffuse
component as well as localized ``radio structures'', such as an internal ring and very compact knots. For simplicity, here we assumed that the synchrotron photons at radio frequencies are all associated with the diffuse electrons in zone 2.

The description of zone 1 is based on our previous results from a detailed study of non-thermal X-ray filaments near the forward shock of
Cas A \citep{ara10}. In that work we used a synchrotron model that also takes into account the effects of diffusion and advection to explain the spectral
evolution across the filaments. The analysis allowed us to determine the SED of the X-ray emitting electrons
in the filaments as well as other properties such as magnetic field and level of turbulence. We found that the electron distributions are similar among the filaments examined. Here, we assumed that the electron populations in the blobs follow a power-law distribution of index $2.6$, in a local magnetic field of $50 \,\mu$G. Note that the adopted value of the magnetic field is significantly lower than those estimated by associating the width of filaments only with synchrotron cooling of radiating electrons \citep[e.g.,][]{vink03}.

\begin{figure}[htp]
\centering
\includegraphics[width=14cm,height=10cm]{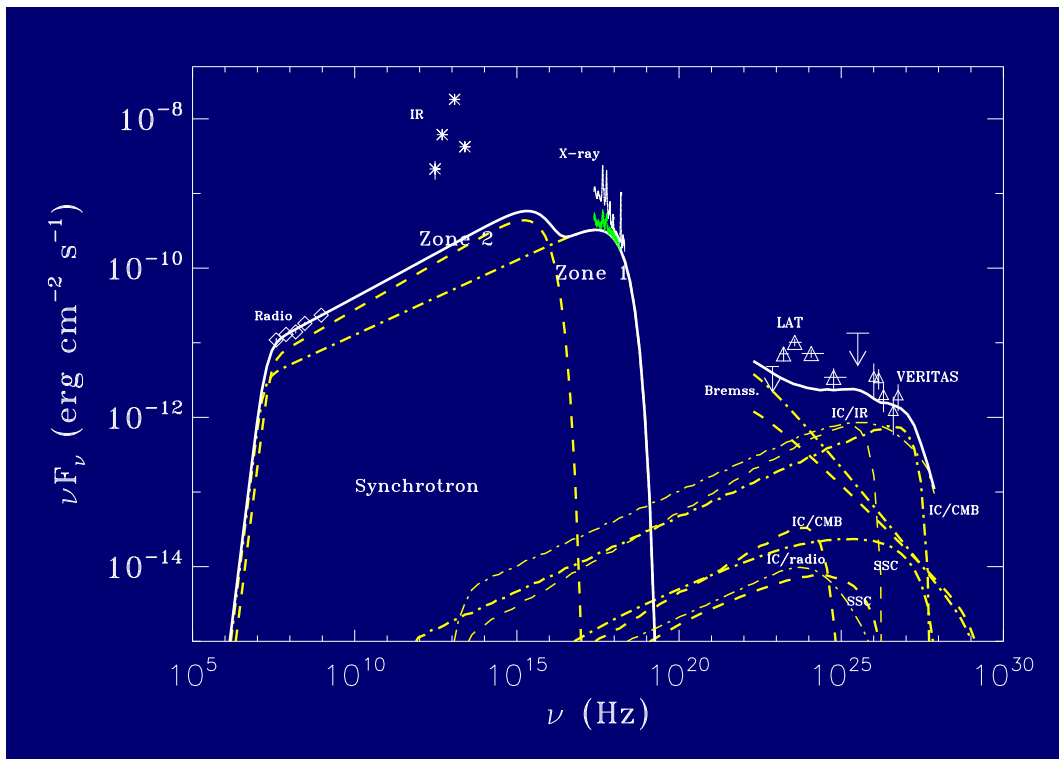}
\caption[SED of Cas A: leptonic contributions]{SED of Cas A and leptonic components. The dashed lines correspond to contributions from zone 2 and the
dash-dotted lines to those from zone 1. Shown in the GeV--TeV bands are bremsstrahlung fluxes (using a proton density for zone 1 of $0.8$ cm$^{-3}$), synchrotron self-Compton
 and IC scattering of CMB (IC/CMB), infrared (IC/IR), and radio (IC/radio) photons. For comparison with the model, we show, separately, the scaled-up X-ray spectrum of a non-thermal filament (in green). Note that the spectrum is ``contaminated'' by thermal emission at low energies (see Araya et al. 2010).
\label{fig-2}}
\end{figure}

For zone 2, we modeled the electron distribution as a power law with a spectral index of 2.54, which reproduces the observed radio spectrum $J_{\nu}\;\propto\; \nu^{-0.77}$ \citep{baa77}. The maximum energy of the electrons is poorly constrained by the data. To assess the maximum contribution from zone 2 to gamma-ray production in Cas A, we pushed the maximum electron energy to as high as possible without causing inconsistency with the X-ray measurement (see Figure~\ref{fig-2}). A mean magnetic field of $\sim300\,\mu$G was used to fit the data. This value is consistent with previous estimates derived from fits to the radio fluxes from the remnant (Atoyan et al. 2000). The overall normalization chosen gives an energy content in the particles that is also consistent with previous results (see Section 4). The high magnetic field could be caused by reverse shock (or secondary shocks) in the interior of the remnant \citep[e.g.,][]{lam01,uchi08}.

Figure \ref{fig-2} also shows the results from leptonic modeling of the SED. The synchrotron emission from zones 1 and 2 is able to account for the non-thermal radio and X-ray fluxes. From the particle distributions, we calculated the number of gamma-ray photons emitted at the source per unit volume per second in the energy range $\epsilon_{\gamma}$ to $\epsilon_{\gamma} + d \epsilon_{\gamma}$ as
\begin{equation}\frac{dn_{\gamma}(\epsilon_{\gamma})}{dt}=\int_0^\infty\,\frac{dn_{\gamma}(T_{e},\epsilon_\gamma)}{dt}\,\left(\frac{4\pi}{v_{e}}\right)\,
\left(\frac{dJ}{dT}\right)_{e}\,dT_{e},\end{equation}
where $(4\pi/v_{e})(dJ/dT)_{e}$ is the number density per unit kinetic energy of electrons ($T_e$) and $dn_\gamma (T_e,\epsilon_\gamma)/dt$ is the emissivity of one electron. The expected photon flux can then be determined for zones 1 and 2 (with volumes $V_1$ and $V_2$, respectively) as
 \begin{equation}\frac{V_{1,2}}{4\pi d_{\mbox{SNR}}^{\,\,\,2}}\, \frac{dn_\gamma^{1,2} (\epsilon_\gamma)}{dt}.\end{equation}

For the two zones, we included bremsstrahlung emission from electron--electron and electron--ion interactions, as well as IC scattering. The production rate of bremsstrahlung photons by an electron interacting with ambient protons, electrons, and helium ions with densities
$n_p$, $n_e$, and $n_{He}$, respectively, is given by
\begin{equation}\frac{dn_\gamma (T_e,\epsilon_\gamma)}{dt} = v_e [(n_p+4n_{He})\,\sigma_{e,p}(T_e,\epsilon_\gamma)+n_e \,\sigma_{e,e}(T_e,\epsilon_\gamma)],\end{equation}
where $\sigma_{e,e}$ and $\sigma_{e,p}$ are the electron--electron and electron--ion cross sections, differential in photon energy and $v_e$ is the
relative speed in bremsstrahlung collisions. The electron--ion cross section depends on the charge as $\sigma_{e,p} \,\, \propto \,\, Z^2$, which explains
the factor $n_p+4n_{He}$ in Equation 2. We considered only interactions of electrons with other electrons, protons, and fully ionized helium. This
implies that $n_e = 1.2\,n_p$. We also assumed an abundance for helium of $n_{He} = 0.1\,n_p$.

The values of $n_p$ used in the calculations of expected bremsstrahlung fluxes can be different for the two zones. For diffuse electrons in zone 2, the density $n_p^{(2)}$ is determined by the average density of protons in the remnant. If we consider uniform distribution of mass in a sphere of radius $R_{\mbox{SNR}}$, then we get a mass for Cas A of $$M_{\mbox{Cas A}} \sim (2.26\, M_{\odot})\frac{n_p}{1\, \mbox{cm}^{-3}}.$$
Therefore, for an estimated remnant mass of $\sim 10\, M_{\odot}$ \cite{wil03}, we get $n_p^{(2)} \sim 4.4\,$cm$^{-3}$. We fixed this value in the
calculations. On the other hand, for the bremsstrahlung flux from electrons in zone 1 the density of ambient protons, $n_p^{(1)}$, cannot exceed $0.8\,$cm$^{-3}$, in order for the model to be compatible with the \emph{Fermi} LAT measurements (see Figure \ref{fig-2}). The fit shown in Figure \ref{fig-2} was based on $n_p^{(1)} = 0.8\,$cm$^{-3}$.

As for IC scattering, we considered various sources of seed photons. The most important one turns out to be infrared photons from the dust in the SNR. The IC scattering of these photons by electrons in zone 2 is calculated under the assumption that the infrared radiation is distributed uniformly across the SNR (which is another gross simplification, as the IR intensity is seen to decrease going outward \cite{dwe87}). On the other hand, in order to fit the TeV fluxes, we found that about $15 \%$ of the total infrared photons reported by the \emph{IRAS} observation should interact with the X-ray-emitting electrons located in zone 1 near the boundary of Cas A. The IC scattering of other seed photons such as the cosmic microwave background (CMB) by electrons in zones 1 and 2, as well as the scattering of radio synchrotron photons from zone 2 by electrons in zone 1, were also included in the calculation. Figure \ref{fig-2} shows that these IC processes are only a minor contributor to the overall gamma-ray production in Cas A.

As also shown in Figure \ref{fig-2}, even with the choice of extreme electron energies for zone 2, the gamma-ray emission from this zone is still less than that from zone 1. It appears, therefore, that leptonic gamma-ray production in Cas A originates mainly near the forward shock.

\subsection{Need for a hadronic contribution}
It is clear from Figure \ref{fig-2} that our leptonic modeling fails to explain the observed gamma-ray emission at GeV energies. An additional component is needed to account for a distinct GeV feature of the SED. The presence of high-energy protons would naturally provide such a spectral component. Collisions between these protons and ambient (cold) protons lead to the production of pions. The decay of neutral pions produces gamma rays.

\begin{figure}[htp]
\centering
\includegraphics[width=14cm,height=10cm]{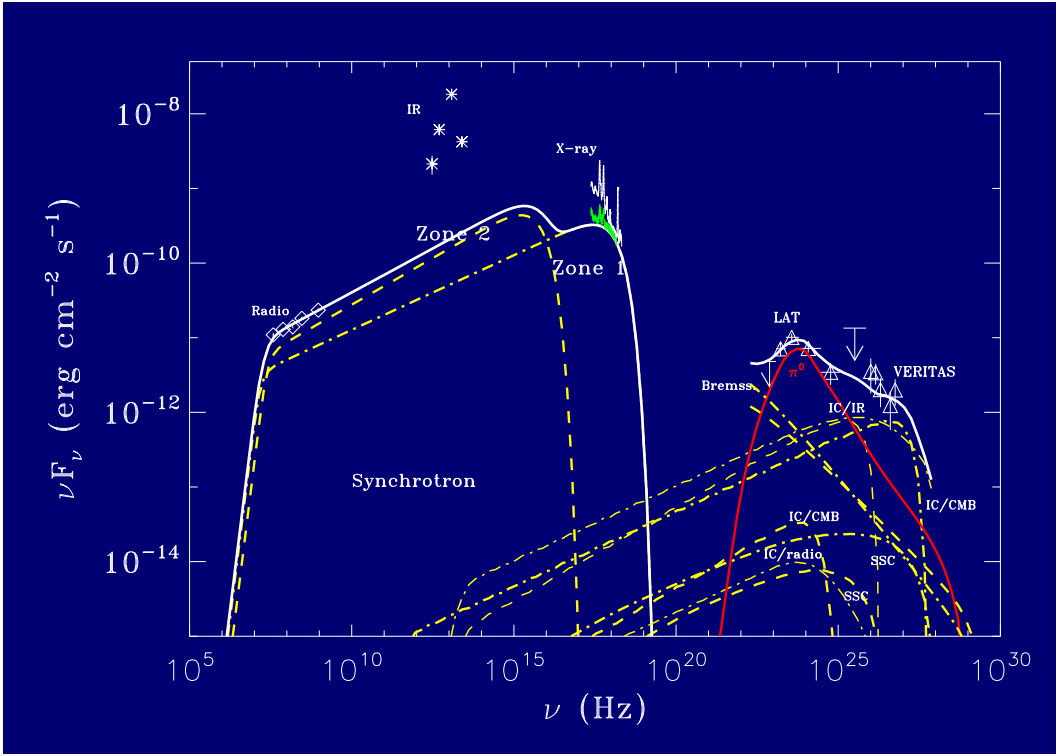}
\caption[SED of Cas A: leptonic and hadronic components]{Modeling of the SED of Cas A with an additional hadronic component. The SED of protons is described by a broken power law in kinetic energy (see the text). For calculating the bremsstrahlung flux from zone 1, we adopted the value $0.5$ cm$^{-3}$ for the ambient proton density, as opposed to $0.8$ cm$^{-3}$ used in Figure \ref{fig-2}.
\label{fig-3}}
\end{figure}

Figure \ref{fig-3} shows the SED of Cas A with an additional component produced by $\pi^0$ decay. The gamma-ray spectrum from hadronic interactions was calculated as in Kamae et al. \cite{kam06}, with a nuclear enhancement factor of 1.9 \citep{mor09}. The data seem to require that the spectral distribution of relativistic protons be a broken power law. In terms of the kinetic energy of the protons ($T_p$), the required distribution has the form $dN/dT_{p} \propto T_p^{-2.1}$ for $T_p < 11$ GeV and $dN/dT_{p} \propto T_p^{-2.7}$ for $T_p > 11$ GeV.

The model shown in Figure \ref{fig-3} was obtained with the parameters in Table \ref{tab-1}. For consistency with the upper limit obtained by the \emph{Fermi} LAT in the 200--500 MeV band, a local density of $n_p^{(1)} = 0.5\,$cm$^{-3}$ was required for the bremsstrahlung emission from zone 1, which is lower than the value used for the model shown in Figure \ref{fig-2} (which has no hadronic contribution).

\section{Discussion}
The \emph{Fermi} satellite, launched on 2008 June 11, has opened a new window of exciting observations in the GeV regime. Hints of hadronic processes in SNRs have been seen. For instance, the \emph{Fermi} LAT spectrum of the middle-aged SNR W51C seems to indicate a dominant $\pi^0$-decay contribution \citep{abd09}. With respect to Cas A, the \emph{Fermi} LAT team discarded the possibility that the GeV emission comes from the compact object near the center of the
remnant on the basis of spectral and timing analysis \citep{abd10b}. They carried out modeling of only the GeV--TeV gamma-ray spectrum in leptonic and hadronic scenarios separately.

In this work, we modeled the radio and X-ray emission from Cas A as synchrotron radiation from two populations of electrons. The radio emission is attributed to diffuse electrons distributed over the whole remnant, while the X-ray emission to electrons localized in about 430 ``blobs'', each with a fixed diameter of $0.32$ pc, which is a typical scale for the width of X-ray filaments near the forward shock. The combined volume of the ``blobs'' is about $1/10$ of the volume of the SNR. The total energy in the electrons is $W_e\sim1.2\times10^{50}\,$erg or about $3\%-6\%$ of the estimated kinetic energy in the explosion of Cas A \cite{lam03}. Bremsstrahlung radiation from the electrons can account for bulk of gamma-ray emission below about 1 GeV, while IC scattering of infrared photons by the electrons seems to be mainly responsible for gamma rays at TeV energies. Although there is considerable degeneracy in modeling the radio to X-ray SED of the source, we found it difficult to account for the GeV spectrum of Cas A with leptons alone.

Adding a $\pi^{0}$ component to the model can explain the distinct feature of the SED between about 1 and 40 GeV. The best match to the data is obtained for protons whose kinetic energies follow a broken power law with indices of $\sim 2.1$ for $T_p < 11$ GeV and $2.7$ for $T_p > 11$ GeV. Spectral steepening of the distribution toward higher energies is required to avoid exceeding the measured fluxes at TeV energies, where IC contribution is high. In our model, IC and $\pi^{0}$ components are comparable between 40 and 500 GeV, where the SED is relatively flat. The total energy in the protons is about $W_p\sim 0.6\times 10^{50}$ erg. However, this value could get higher if a lower percentage of infrared emission is assumed to be interacting with the X-ray filaments (which would make IC contribution less important).

Our model contains several simplifications. For instance, we neglected the spatial inhomogeneities in the radio brightness distribution, which point at the existence of additional electron populations (Atoyan et al. 2000). However, as calculated by Atoyan et al. (2000), the expected bremsstrahlung flux from these electrons is considerably less than that of the diffuse component at GeV energies due to their steeper distributions. It can be argued similarly that the expected emission from IC scattering of ambient photons off these electrons is low, which justifies neglecting their contribution to the SED. We also assumed a uniform proton density inside Cas A, which is not realistic. We will investigate the effects of a more realistic density profile for young SNRs in a future work.

Vink and Laming (2003) also modeled the emission from Cas A with two populations of electrons. They adopted a higher magnetic field ($80 $-$160 \,\mu$G) near the shock, which is obtained when comparing a typical width of an X-ray filament with the synchrotron cooling length of a typical high-energy electron \citep[see also][]{par06}. In our model we decided to use a lower value ($50\,\mu$G) derived by Araya et al. (2010) who took into account the effects of diffusion, in addition to those of synchrotron loss and advection. With their model, Vink and Laming (2003) predicted a gamma-ray flux from IC scattering of ambient photons that lies below the HEGRA observation at TeV energies which, they argued, might be a hint to a pion decay origin of the radiation. In our model, most of the emission at TeV energies is attributed to IC scattering of infrared photons, which is a consequence of adopting a lower magnetic field at the shock front. It should be stressed that the data points at GeV energies, which were not available to previous studies, hold the key to revealing the presence of hadronic processes in Cas A.

\begin{figure}[htp]
\centering
\includegraphics[width=14cm,height=10cm]{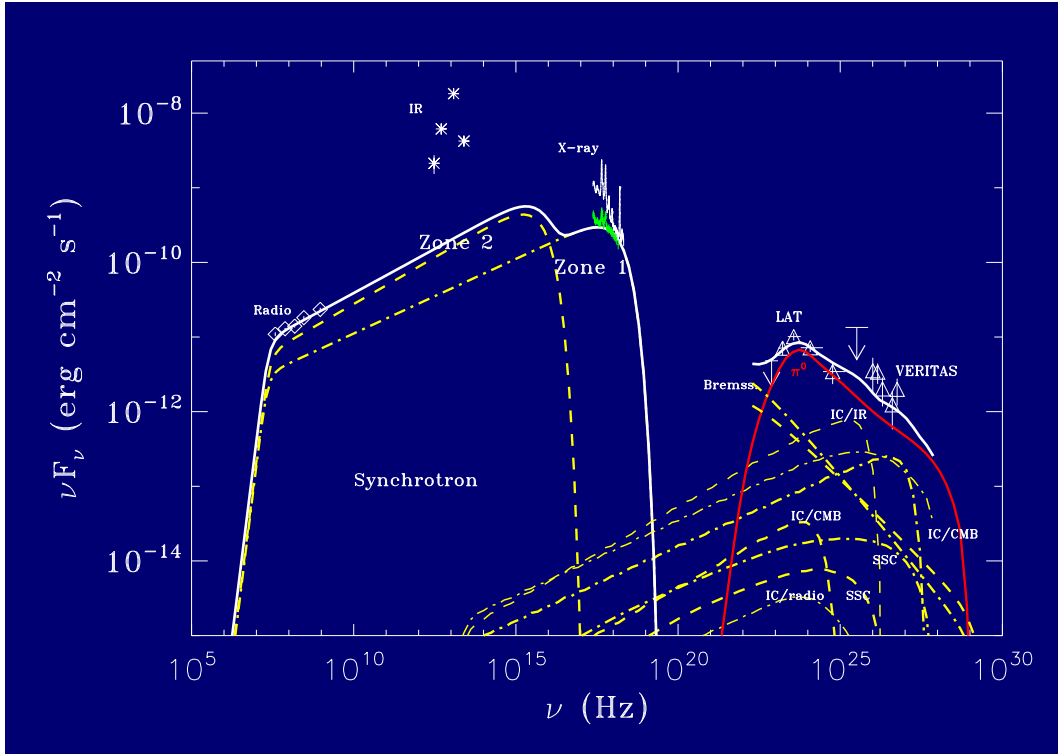}
\caption[SED of Cas A: different parameters]{Same as Figure \ref{fig-3}, but with a magnetic field of $90\,\mu$G and an ambient proton density of $1.5\,$cm$^{-3}$ in zone 1.
\label{fig-4}}
\end{figure}

Although we believe that the lower value of the magnetic field  that we adopted is more reliable, the measurement of the field is still quite uncertain. We also ran models with higher magnetic field.
As an example, Figure \ref{fig-4} shows a reasonable fit to the data with $B=90\,\mu$G in zone 1. The higher field causes a reduction in the IC contribution, which is compensated by a harder proton distribution (with a spectral index of $2.4$ for kinetic energies $T_p > 8$ GeV). We also had to use a higher ambient proton density ($n_p^{(1)} = 1.5\,$cm$^{-3}$) and a lower volume for zone 1 ($\sim V_{\mbox{SNR}}/30$). For $B > 90\,\mu$G in zone 1, we found it difficult to explain the observed high-energy SED with our model.

In summary, we were unable to account for the observed gamma-ray fluxes of Cas A at GeV energies with leptons alone in our model. The GeV ``excess'' can be well explained by additional gamma rays from the decay of neutral pions, which may be produced in the interactions between relativistic hadrons and dense matter in the remnant. This constitutes evidence for cosmic ray acceleration in Cas A.

\acknowledgments
We thank T. Kamae for kindly providing us with his code for calculating the cross sections and gamma-ray fluxes in \emph{pp} interactions, and S. Funk for providing his Fermi LAT fluxes of Cas A in electronic format for comparison. We also thank M. Lyutikov and D. Lomiashvili for useful discussions. This research has made use of data obtained from the \emph{Chandra} Data Archive and the \emph{Chandra} Source Catalog, and software provided by the \emph{Chandra} X-ray Center (CXC) in the application package CIAO. This work has also made use of NASA's Astrophysical Data System. We gratefully acknowledge the financial support from NASA and Purdue University.

\newpage
\begin{table}
\begin{center}
\caption{Model Parameters \label{tab-1}}
\begin{tabular}{|c|c|c|c|}
\tableline\tableline
&\multicolumn{2}{c|}{Leptonic Component}&\multicolumn{1}{c|}{Hadronic Component}\\ \cline{2-3}
\raisebox{0.1in}[0in][0in]{Parameter} &\multicolumn{1}{c}{Zone 1} & \multicolumn{1}{c|}{Zone 2} & \multicolumn{1}{c|}{ }\\
\tableline
Magnetic field ($\mu$G) & 50 & 300 & - \\
Ambient proton density (cm$^{-3}$) & 0.5 & 4.4 & 4.4 \\
Minimum Lorentz factor & $1$ & $1$ & $1.5$ \\
Maximum Lorentz factor & $7\times10^{7}$ & $2\times10^{6}$ & $5.5\times10^{5}$ \\
Volume &$\sim \frac{1}{10}V_{\mbox{SNR}}$& $V_{\mbox{SNR}}$ & $V_{\mbox{SNR}}$ \\
\multirow{2}{*}{Spectral index\tablenotemark{*}} & 2.6 & 2.54 & 2.1, $\,\,\,T_p <11$ GeV \\ & & & 2.7, $\,\,\,T_p >11$ GeV \\
\tableline
Total energy (erg) & \multicolumn{2}{|c|}{$1.2\,\times\,10^{50}\,$} & $6.3\,\times\,10^{49}\,$ \\
\tableline
\end{tabular}
\end{center}
\end{table}

\clearpage

\end{document}